\documentclass[epjST]{svjour}
\usepackage{graphicx}
\usepackage{amssymb} %%necessary for math symbols
\usepackage[utf8]{inputenc}
\usepackage{xcolor}

\AtBeginDocument{%
  \paperwidth=\dimexpr
    1in + \oddsidemargin
    + \textwidth
    + 1in + \oddsidemargin
  \relax
  \paperheight=\dimexpr
    1in + \topmargin
    + \headheight + \headsep
    + \textheight
    + 1in + \topmargin
  \relax
  \usepackage[pass]{geometry}\relax
}

\begin{document}
\title{Dynamical trapping in the area-preserving H\'enon map}
\author{V. M. de Oliveira\inst{1}\fnmsep\thanks{\email{vitormo@if.usp.br}} \and D. Ciro\inst{2} \and I. L. Caldas\inst{1} }
\institute{Instituto de F\'isica, Universidade de S\~ao Paulo, S\~ao Paulo, SP, Brazil \and Instituto de Astronomia, Geof\'isica e Ci\^encias Atmosf\'ericas, Universidade de S\~ao Paulo, S\~ao Paulo, SP, Brazil}
\abstract{
Stickiness is a well known phenomenon in which chaotic orbits expend an expressive amount of time in specific regions of the chaotic sea.
%, specially occurring around resonance island chains. 
This phenomenon 
%can be induced by invariant manifolds and it
becomes important when dealing with area-preserving open systems because, in this case, it leads to a temporary trapping of orbits in certain regions of phase space.
In this work, we propose that the different scenarios of dynamical trapping can be explained by analyzing the crossings between invariant manifolds. In order to corroborate this assertion, we use an adaptive refinement procedure to approximately obtain the sets of homoclinic and heteroclinic intersections for the area-preserving Hénon map, an archetype of open systems, for a generic parameter interval.
%The homoclinic set is formed by the intersections of the manifolds belonging to a periodic saddle of period 12 and the heteroclinic one is composed by the intersections between the manifolds associated with the aforementioned periodic orbit and the two fixed saddles presented in the system.
%We chose a parameter interval where different scenarios of dynamical trapping are present, which is reflected in phase space as how uniform is the chaotic sea. To quantify this phenomenon, we calculate the transit time for orbits in the chaotic region, i.e., the time it takes to cross the system.
%We observe that the geometry of the chaotic region in phase space is strongly influenced by manifolds associated of fixed saddles while the orbits with long transit times are mostly influenced by the manifolds of a periodic saddle of period 12 around a structure of regular solutions.
We show that these sets have very different statistical properties 
%regarding their distribution
%in phase space
when the system is highly influenced by dynamical trapping, whereas they present similar properties when stickiness is almost absent.
We explain these different scenarios by taking into consideration various effects that occur simultaneously in the system,
%such as changes in the accessible region for the chaotic orbits and preferred concentration of crossings,
all of which are connected with the topology of the invariant manifolds.
%In this work we study the \change{evolution} of homoclinic and heteroclinic connections of the area-preserving Hénon map as an archetype of open conservative systems. The map has two saddle points and their invariant manifolds are used to characterize the access region and the general geometrical features of the chaotic portion of phase space, while the manifolds of an internal periodic saddle are related to the existence of long transit time orbits around a detaching island chain in the chaotic domain. It is shown that the reduction of persistent orbits during the island chain detachment from the regular region is due to the redistribution of homoclinic points associated with the periodic saddle, along with the proliferation of heteroclinic connections between the periodic saddle and the fixed points at the entry and exit regions, leading to a more uniform chaotic sea.
%In general, invariant manifolds of periodic saddles in the chaotic region determine the geometry of the income and exit sets in systems with chaotic scattering, and the manifolds topology in the chaotic domain have an important influence on the transit time of nearby orbits. However, the manifolds distribution in phase space requires a comprehensive knowledge of the invariant sets. In this work, we study the influence of homoclinic and heteroclinic connections in the transit time of the area-preserving Hénon map, an archetype of open conservative systems, to show how the transit time changes with the manifold connections transformation as a control parameter is modified.
} %end of abstract
\maketitle
\section{Introduction} \label{sec:intro}
Chaotic solutions of unbounded area-preserving maps usually consist of an incoming regular path, a transitory irregular motion and a regular exit path. In simple situations the irregular motion occurs within a localized region of phase space and individual orbits do not access the whole chaotic domain, i.e. the chaotic orbits are not transitive \cite{Wiggins2003}. During its irregular motion each orbit spends a finite amount of time wandering in a definite sub-region of the chaotic domain which is determined by its income path.

The average transit time of the irregular orbits can be related to the measure of the accessible region by an incoming set~\cite{Meiss1997}, but the details of the transit time distribution in phase space are more subtle and require comprehensive knowledge of the invariant sets in phase space. In general, invariant manifolds of periodic saddles in the chaotic region determine the geometry of the income and exit sets in systems with chaotic scattering~\cite{Petrisor2003,Tel2006}, and the manifolds topology in the chaotic domain have an important influence on the transit time of nearby orbits~\cite{Contopoulos2010}.

%\change{(Ibere) The dynamical trapping is associated to the stickiness observed whenever the manifolds crossings become significant around the invariant islands. However, this article deals with the origin of these effects, i.e., the manifold distribution in phase space...}
%
%\change{(Vitor) Even though stickiness is the reason behind the studied effect, namely dynamical trapping, it is not the focus of the manuscript.}

Dynamical trapping occurs when the transit time increases in certain regions of phase space. This behavior is associated to the stickiness observed whenever the manifolds crossings become significant around the invariant islands, which is a well known phenomenon in conservative systems \cite{Zaslavsky1984,Altmann2006}. In this paper, we are interested in the origin of these effects, i.e., the manifold distribution in phase space and the influence of homoclinic and heteroclinic connections in the transit time of area-preserving maps.

%In this work, we are interested in studying the influence of homoclinic and heteroclinic connections in the transit time of area-preserving maps.

We illustrate our numerical approach by means of the area-preserving Hénon map, which is algebraically simple and has all the desired features of an open system, including a set of homoclinic orbits in the chaotic sea~\cite{Devaney1984,Kirchgraber2006}. In particular, we study the interaction between three important unstable periodic solutions in a parameter range, two period-1 saddles located at the enter and exit portions of the irregular region and one period-12 saddle associated with a regular island immersed in the chaotic sea. The geometry of the accessible region in phase space is strongly influenced by the period-1 manifolds while the orbits with long transit times are mostly influenced by the manifolds of the period-12 saddle. We show this fact by studying the intersections between the period-1 and period-12 manifolds, which establish the paths for dynamical trapping in the chaotic sea, so that their distribution determine the presence of the most persistent orbits in phase space.

This paper is organized as follows. In Sec.~\ref{sec:map} we present the area-preserving H\'enon map and its characteristic dynamical features. In Sec.~\ref{sec:inv_man} we trace and discuss on the invariant manifolds of the most relevant saddles of the map. In Sec.~\ref{sec:int_ana} we study the distribution of the homoclinic and heteroclinic connections in phase space and, finally, we present our conclusions in Sec.~\ref{sec:conc}.

%\cite{Meiss1997} Meiss article: defines transit time and average exit time. Applies them for the area-preserving Henon map. Relates them to specific measures and available areas in phase space.

%\cite{Petrisor2003} Petrisor article: uses the term ``chaotic scattering" associated with the area-preserving Henon map. Shows that the entry and exit sets follows the manifolds of the fixed saddles of the system.

%\cite{Devaney1984} Devaney article: develops a geometric method from proving the existence of homoclinic orbits and uses the area-preserving Henon map as an example.

%\cite{Ritter1987} Ritter article: uses normal formal to compute homoclinic points and UPO in area-preserving maps. Applies it to a different Henon map, but Alfredo Osorio de Almeida is a co-author of the paper.

\section{The area-preserving H\'enon map} \label{sec:map}

The H\'enon map was first proposed as a simple mapping to describe the dynamical properties of the Lorenz system, which models atmospheric phenomena \cite{Henon1976}. Since then, it has become a hallmark for studying dynamical properties on two-dimensional discrete systems. In this work, we are interested in an area-preserving and orientation-reversing version of the map, which is given by~\cite{Alligood1996}:

\begin{equation}
    \begin{array}{cl}
    x_ {n+1}= & a-x_n^2+y_n,\\
    y_{n+1}= & x_n.
    \end{array}
\label{eq:henon}
\end{equation}

\begin{figure}[h]
    \centering
    \includegraphics[width=0.94\textwidth]{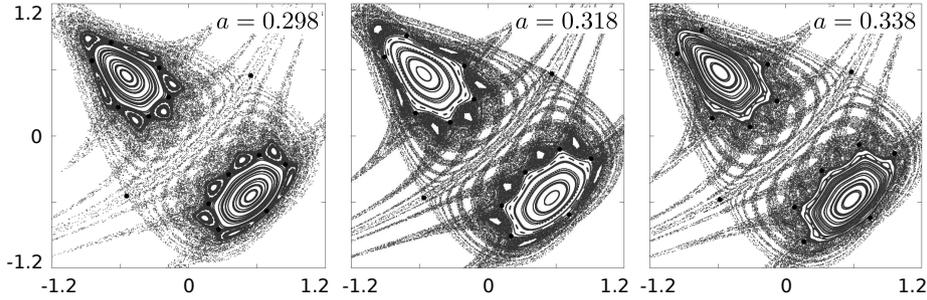}
    \caption{Phase space ($x\times y$) of the area-preserving H\'enon map with orbits entering and exiting the chaotic domain along with some invariant orbits. The black circles indicate two fixed saddles and a periodic saddle of period 12 which will be important latter on.}
    \label{fig:phase_space}
\end{figure}

In Fig. \ref{fig:phase_space} we show a collection of regular and irregular orbits of the map for $a=0.298$, $a=0.318$ and $a=0.338$. In this range we observe the presence of a period-12 resonant island in contact with the open chaos and surrounding a regular region containing a period-2 center.

\begin{figure}[h]
    \centering
    \includegraphics[width=0.8\textwidth]{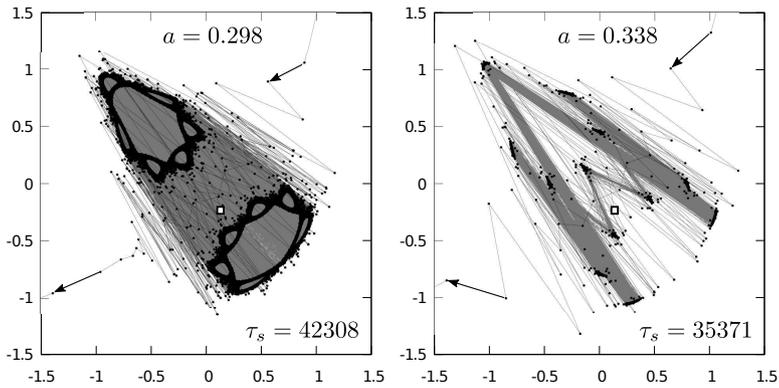}
    \caption{The most persistent orbit in phase space ($x\times y$) with a point inside the marked box for two control parameters. The transit times $\tau_s$ are similar in both cases, but for $a=0.298$ the orbit access a wider region due to the size of the regular islands.}
    \label{fig:max_orbits}
\end{figure}

As was mentioned in the introduction, the area-preserving H\'enon map is an open dynamical system, and, consequently, its irregular solutions are not transitive in the chaotic region of phase space.
In Fig.~\ref{fig:max_orbits} we illustrate this fact by tracing the most persistent chaotic orbit for two control parameters. These orbits are chosen as to have a point inside a small arbitrary region in the chaotic sea (marked box in Fig.~\ref{fig:max_orbits}).
In both cases, the orbit arrives from $+\infty$, and spends a long time in the chaotic region before leaving to $-\infty$, without returning. During the visit, the solutions do not access all the chaotic sea, but mostly visit very localized regions around the period-12 invariant subset, therefore, they are irregular but not transitive. The most persistent solution was chosen to emphasize that even the orbit with the longest transit time does not cover the full chaotic sea, and its transit time value is influenced by the period-12 island chain, even though the arbitrary box is not geometrically close to it.

\begin{figure}[h]
    \centering
    \includegraphics[width=0.98\textwidth]{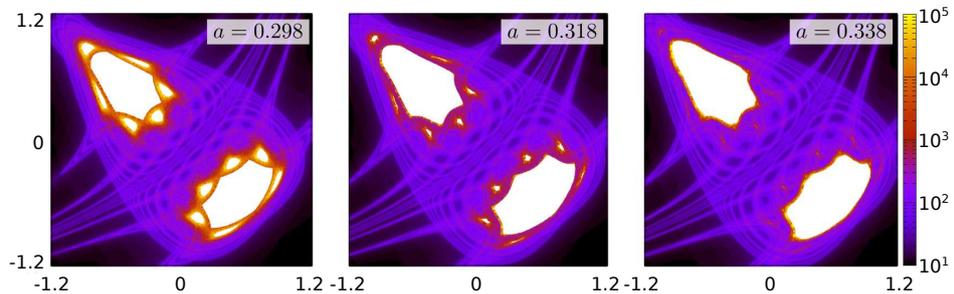}
    \caption{(Color online) Transit time distribution for a grid of initial conditions in phase space ($x\times y$) for different values of $a$. The orbits with larger transit times (yellow/light grey) are concentrated around the period-12 island chain.}
    \label{fig:profile}
\end{figure}

In order to understand the structure of the open chaos, we calculate the transit time distribution in a region containing the periodic solutions of interest. In Fig. \ref{fig:profile} we show the results for a $1000\times 1000$ grid of initial conditions under three different control parameter values. Each orbit was iterated with the Hen\'on map and its inverse until it reached a large box containing the depicted domain. The combined number of iterations corresponds to the transit time.

From the density plots we observe that the most persistent orbits are organized around the period-12 island, in agreement with the observations in Fig.~\ref{fig:max_orbits}. As the control parameter increases the resonant islands become smaller and the number of orbits with long transit times is reduced. Consequently, the transit times become more uniform for larger $a$, leading to changes in the characteristic transit time of the system and the dominant patterns of chaotic behavior.

The light regions in Fig. \ref{fig:profile} clearly converge to the high-persistence region around the islands, and follow the high density pattern of the solutions in Fig.~\ref{fig:phase_space}. As is expected, darker regions in the transit time correspond to the less visited regions in Fig.~\ref{fig:phase_space}, indicating an underlying geometrical structure that separates the orbits with short and long survival times. We explore more of these geometric features in the next section.

\section{Invariant manifolds}
\label{sec:inv_man}

The configuration of the invariant manifolds of fixed or periodic saddles gives the underlying structure of the system dynamics. These curves are formed by the continuous collection of orbits that converge asymptotically to the saddle by forward (stable manifold) or backwards (unstable manifold) iteration.

\begin{figure}[h]
    \centering
    \includegraphics[width=0.94\textwidth]{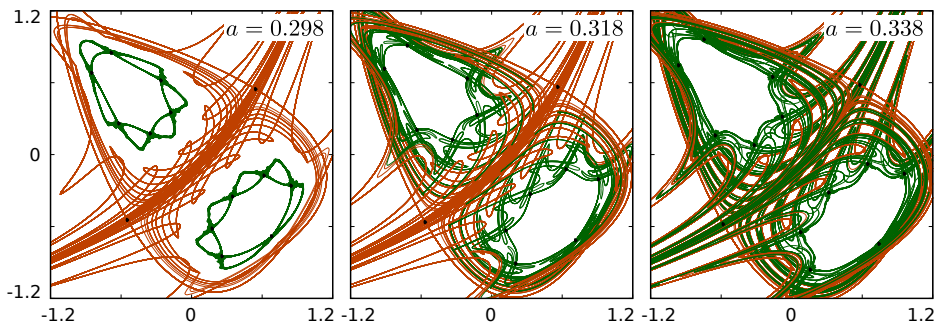}
    \caption{(Color online) Invariant manifolds in phase space ($x\times y$) of the fixed points (orange/dark grey) and the period-12 orbit (green/light grey) for different values of $a$. As the control parameter increases the fixed points and periodic saddle manifolds become more entangled.}
    \label{fig:manifolds}
\end{figure}

The area-preserving H\'enon map, eq. (\ref{eq:henon}), has two fixed saddles which are located at $x=y=\pm\sqrt{a}$. For the range of parameters analyzed, namely, $a\in[0.298,0.338]$, it has also a period-2 center which is surrounded by a period-12 saddle (see Fig. \ref{fig:phase_space}). In Fig. \ref{fig:manifolds}, we present the invariant manifolds of these saddles, which were calculated using an efficient tracing method developed for planar maps \cite{Hobson1993,Ciro2018}.

Before analyzing the manifolds interaction an important observation is in order: two orbits separated by a manifold must remain separated by that manifold at all times. Therefore, the manifolds of the period-1 saddles separate orbits in a set that access the chaotic region and a set that circumvent that region and move to infinity. This is clearly observed in Fig.~\ref{fig:phase_space}, notice that the chaotic orbits are packed in phase space by the manifolds depicted in Fig.~\ref{fig:manifolds}. Consequently, the size of the chaotic region where the interesting dynamics takes place is determined by the area delimited by these curves in phase space, defining a forbidden region where the scattering orbits cannot access.

For $a=0.298$, the manifolds of the fixed saddles are separated from the manifolds of the period-12 saddle. Even though these structures do cross eventually (due to the fact that they are immersed in the same chaotic region), this occur only after a large number of lobes were formed by both manifolds. For $a=0.318$, the crossings start sooner and the manifolds can no longer be considered independent, as curves with the same type of stability must be parallel (in the sense of never crossing) and this leads to a similar behavior of the green and orange manifolds. Finally, for $a=0.338$, this behavior becomes more accented, and the manifolds entangle completely, leading to long excursions of the period-12 manifold to the open chaos as it follows the period-1 manifolds across the phase space.

We can now compare Figs. \ref{fig:profile} and \ref{fig:manifolds} to understand the relation between the invariant manifolds crossings and the chaotic behavior of the system. For $a=0.298$, the invariant manifolds of the period-1 saddles present homoclinic crossings with themselves and heteroclinic crossings between them, while the manifolds of the period-12 island mostly present homoclinic crossings, which are highly localized in phase space. From this scenario we can infer that chaotic orbits near the crossings of the period-1 saddles move in a wider area and wander between the neighborhood of these fixed points, while orbits near the crossings of the period-12 saddle spend most time in a localized region with limited interaction with the open chaos. As mentioned before, heteroclinic crossings between the period-1 and period-12 manifolds do occur for very long tracings of the manifolds, emphasizing the fact that \emph{few} orbits from the open chaos do access the chaos around the islands and, in doing so, they must spend long times there, therefore, becoming temporarily ``trapped" by a manifold induced stickiness~\cite{Contopoulos2010}. These observations are in agreement with the high values of survival time observed around the island chain for $a=0.298$ in Fig.~\ref{fig:profile}.

For $a=0.318$, there is an increase of heteroclinic crossings between the period-1 and period-12 saddles and the previously localized homoclinic crossings become more sparse. Consequently, the number of orbits that move around the island chain increases but their permanence is reduced due to the stronger interaction with the manifolds of the fixed saddles. Finally, for $a=0.338$, all the manifolds are highly correlated. Hence, the mechanism that confines the manifolds of the period-12 saddle in smaller regions of phase space is not available and the interaction between the manifolds from different saddles becomes dominant. With this, the homoclinic crossings of the period-12 island are not so localized, leading to a more uniform distribution of orbits in the chaotic region and to a lower number of orbits with high survival times.

In short the manifolds of the fixed saddles are responsible for the transport of orbits from the outer to the inner region, and vice-versa, whilst the ones associated with the resonant island chain are responsible from temporarily trapping the orbits in the inner region. The heteroclinic crossings, by their turn, connect both phenomena, and the increase of the control parameter reduces the complexity of the system as the homoclinic and heteroclinic crossings distribute in a more similar fashion, leading to a more uniform transit time distribution in phase space.

To better elucidate the different behaviors of the system for different control parameters, in the following section we analyze the distribution of crossings in phase space and their relation to the structure of the open chaos.

\section{Intersections analysis} \label{sec:int_ana}
We now proceed to compare the set of homoclinic crossings of the period-12 unstable periodic orbit near the last invariant curve, with the set of heteroclinic crossings between the periodic saddle manifolds and the ones associated with the period-1 saddles of the map. The sets were calculated to a given accuracy by means of a refinement procedure that generated an improved group of points around the best performing points of a previous group. The performance of a point was measured in terms of the shortest distance to the chosen saddles achieved by its associated finite forward and backwards orbit. The refinement was then carried out up to a given tolerance on the orbit proximity to said saddles.

\begin{figure}[h]
    \centering
    \includegraphics[width=0.95\textwidth]{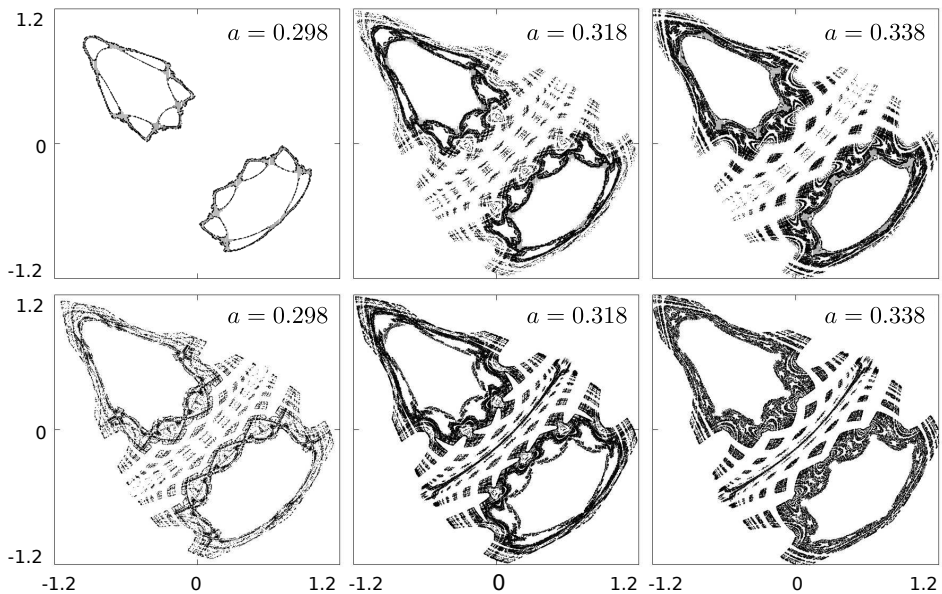}
    \caption{Approximated homoclinic (top) and heteroclinic (bottom) points in phase space ($x\times y$) for three different parameters of the area-preserving H\'enon map. As the control parameter increases the homoclinic and heteroclinic crossings become more correlated.}
    \label{fig:crossings}
\end{figure}

In Fig.~\ref{fig:crossings} we depict the homoclinic and heteroclinic sets in phase space for different values of the control parameter $a$. As the parameter increases, the region occupied by the homoclinic points grows rapidly, from the vicinity of the period-12 island to the open chaos region, and, at the same time, the heteroclinic points become more dense around the island chain, enhancing the connection between it and the rest of the chaotic sea.
The clear correlation that arises between the homoclinic and heteroclinic sets is due to the parallel nature of the manifolds that arrive at the period-12 and period-1 saddles and have the same stability (vide Fig. \ref{fig:manifolds}). Such curves are parallel in the sense that they are not allowed to cross, and this causes two close manifolds to fold in the same fashion, generating locally similar patterns and, consequently, locally similar crossings with their counterpart of opposite stability. To understand this we need to quantify the distribution of crossings in phase space.

\begin{figure}[h]
    \centering
    \includegraphics[width=0.65\textwidth]{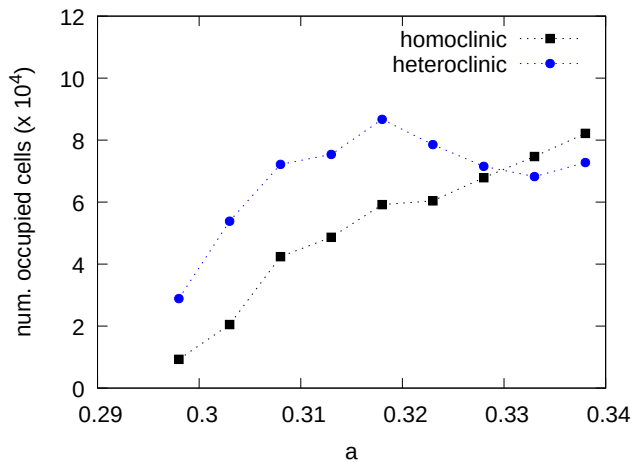}
    \caption{Number of occupied cells in phase space as a function of $a$. The advance of the forbidden region into the chaotic sea leads to the reduction of available cells, which mostly affects the heteroclinic points.}
    \label{fig:num_occ}
\end{figure}

To describe the evolution of the homoclinic and heteroclinic sets we define a $1000\times 1000$ grid on the region of interest in phase space and study the statistics of the occupation numbers on the cells as the parameter changes. First, we estimate the total area of phase space occupied by the homoclinic and heteroclinic points by counting the number of occupied cells, as shown in Fig. \ref{fig:num_occ}.
 As expected, the number of cells containing homoclinic points in phase space grows monotonically with the control parameter, in agreement with the observations in Fig.~\ref{fig:crossings}. However, the same trend is not observed for the heteroclinic set, which is influenced by the fast-exit regions defined by the manifolds of the period-1 saddles. As the parameter grows, the manifold lobes deepen into the chaotic domain reducing the available space for both sets. Since the heteroclinic set has more limited access to the internal region of the islands, this gives a small advantage to the homoclinics, which end up occupying more cells in space.

\begin{figure}[h]
    \centering
    \includegraphics[width=0.65\textwidth]{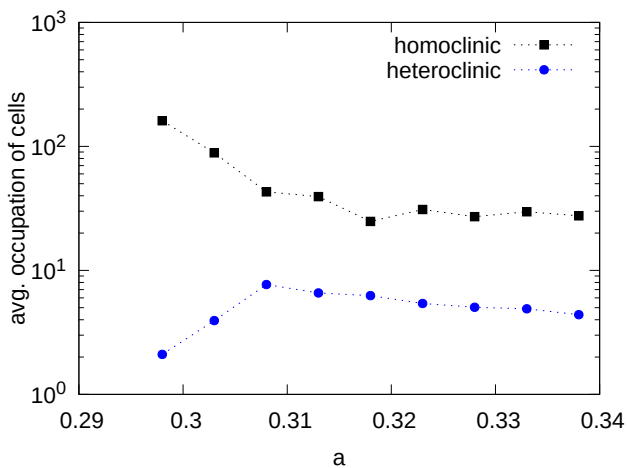}
    \caption{Average cell occupation for the homoclinic and heteroclinic points. After a redistribution of homoclinic points and a exponential increase of heteroclinic points both sets give the same trend.}
    \label{fig:avg_occ}
\end{figure}

In order to quantify the density of the homoclinic and heteroclinic sets in phase space we measure the average occupation number inside the cells as a function of $a$, which is presented in Fig. \ref{fig:avg_occ}.
Homoclinic points are always far more abundant than the heteroclinic ones because the crossings that define them exist before the merging of the local chaos (about the island chain) and the open chaos (between the period-1 saddles). Hence, heteroclinic points must be created at a large rate as we increase the control parameter, resulting in an exponential growth of the size of the set as is observed in the average occupation numbers between $a=0.298$ and $a=0.308$. On the other hand, there is an exponential decay in the average occupation of homoclinic points for the same parameter range due to the rapid increase in their used area in phase space, indicating that their absolute number is not growing, but only redistributing. After $a=0.308$ the average occupation of both sets decay slowly and in the same fashion, indicating a tendency to uniform their distribution in phase space. In particular, after this parameter value, the heteroclinic crossings reduce in average cell concentration and in number of occupied cells as well. To understand how, we need to further analyze the statistics of cell occupations.

\begin{figure}[h]
    \centering
    \includegraphics[width=0.65\textwidth]{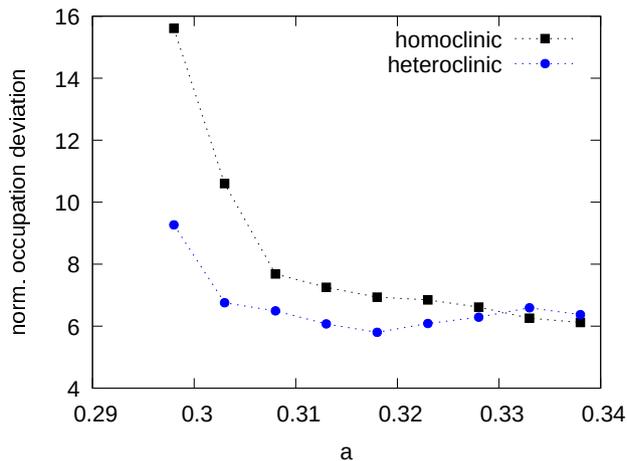}
    \caption{Normalized deviation of the cell occupation for the homoclinic and heteroclinic points. The advance of the forbidden region causes an apparent increase for the heteroclinic points.}
    \label{fig:var_occ}
\end{figure}

We consider the standard deviation of the occupation number $\sigma_n$ in order to characterize the uniformity of the occupation in the cells. This measure is sensitive to the total number of crossings and, therefore, return very different values for the heteroclinic and homoclinic sets. For the sake of comparing both sets in the same footing, we consider a normalized version of this quantity $\bar{\sigma} = \sigma_n/\bar{n}$, where $\bar{n}$ is the average occupation number. The normalized occupation deviation then characterizes the dispersion of the occupation numbers in units of the average occupation and is shown in Fig. \ref{fig:var_occ} as a function of $a$.
As observed in Fig.~\ref{fig:crossings} both homoclinic and heteroclinic crossings redistribute in a more uniform fashion in the allowed region as we increase the control parameter, and this is consistent with initial trends observed in Fig.~\ref{fig:var_occ}. Although the homoclinic points continue to distribute more uniformly, the heteroclinics present a rebound in the dispersion after $a=0.318$. This coincides with the decrease in size of the chaotic region (which reduces the amount of occupied cells in Fig.~\ref{fig:num_occ}) combined with the decrease in the average occupation number, leading to an apparent growth in the normalized deviation. Another clear aspect of Fig.~\ref{fig:var_occ} is that the initial cell occupation of homoclinic points is far more disperse than that of the heteroclinic ones, indicating that their cells are occupied in a more diverse fashion, with frequent high and low concentrations, while the heteroclinic points are initially more uniformly distributed in their respective cells.

At $a \approx 0.33$ we can observe a crossing between the curves associated with the homoclinic and heteroclinic sets in Fig.~\ref{fig:num_occ} and \ref{fig:var_occ}. At this parameter value, the homoclinic set surpasses the number of occupied cells related to the heteroclinic set indicating that a larger area of phase space contains homoclinic points, but not necessarily that there is a higher number of crossings of some kind. Although we can say that the phase space goes from being predominantly heteroclinic to homoclinic, this interchange does not constitute a transition in a dynamical sense, as the system evolution in not too sensitive to the occupied area but rather to the amount of homoclinic and heteroclinic points.

The fact that both sets present similar dispersion at the end of the parameter range indicates that the cell occupations are distributed in a similar fashion in phase space, though very different absolute numbers are observed in Fig.~\ref{fig:avg_occ}. This also coincides with the increase in the manifold correlation of the period-1 and period-12 saddles, as can be seen in Fig.~\ref{fig:manifolds}, which induces similar spatial properties for the homoclinic and heteroclinic sets.

\section{Conclusions}
\label{sec:conc}

In this work, we have studied some general features of the chaotic dynamics of conservative open systems with the aid of a simple archetypal system, the area-preserving Hénon map. By calculating the transit time distribution in phase space, we observed the emergence of different persistence scenarios in the scattering region for a generic parameter interval containing the detachment and almost destruction of a regular resonant island chain. In order to explain such different scenarios, we studied the changes in two sets of points which were defined by the invariant manifolds of two unstable fixed points, located at the entry and exit regions of the chaotic domain, and a periodic saddle, located near a resonant island chain that was immersed in the open chaos. The first set analyzed was formed by the homoclinic connections between the manifolds of the periodic saddle, while the second one comprehended the heteroclinic connections between those manifolds and the ones associated with the fixed saddles. Both sets were obtained approximately by an adaptive refinement procedure.

The first control parameter analyzed was just after the island chain detachment, where the properties of the homoclinic and heteroclinic intersections were quite different. In this scenario the homoclinic points were concentrated around the periodic saddle while the homoclinic/heteroclinic points of the unstable fixed points occupied a relatively large region of phase space. However, heteroclinic connections between the periodic saddle and fixed points were scarce, leading to a layered chaos with longer transit times around the islands. As the control parameter increased, the homoclinic and heteroclinic connections occupied more space, and the correlation between their locations became more explicit, as expected from the manifold calculations.

As the island chain got more detached from the regular solutions, the initial proliferation of heteroclinic points lead to an exponential growth in their density in phase space, while the simultaneous redistribution of homoclinic points lead to an exponential decay in theirs. The heteroclinic points then reached the limit of the accessible region and both densities started to drop slowly and in the same manner, leading to a more uniform distribution of points for both sets.

In summary, it was observed that longer transit times were associated with orbits around the homoclinic connections of the periodic saddles, while the unstable fixed points played a more structural role, delimiting the open chaotic region. As the regular island chain detached, the sticky behavior was reduced by the heteroclinic connection between the fixed points and the periodic saddles and the homoclinic and heteroclinic sets became more similar as their manifolds got more entangled. These observations must hold qualitatively for different parameter ranges, provided that there are resonant islands getting detached from the regular region of phase space and, consequently, a suitable periodic orbit can be chosen.

\section*{Acknowledgement}
\label{sec:ack}
This study was financed in part by the Coordena\c c\~ao de Aperfeiçoamento de Pessoal de Nível Superior - Brasil (CAPES) - Finance Code 001 and the São Paulo Research Foundation (FAPESP, Brazil), under Grant No.  2018/03211-6.

\section*{Author contributions}
\label{sec:aut}
The main ideas behind this work are attributed to all authors. V.M.O. and D.C. performed the numerical simulations and analysis. I.L.C. contributed with results interpretation and manuscript revision.

\end{document}